\documentclass[prl,aps,twocolumn]{revtex4}
\usepackage{graphicx} 
\def\gr{$\gamma$-ray}
\usepackage{color}

\begin{document}
\title{PeV extended emission around Galactic PeVatrons and echos of their flares}
\author{Andrii Neronov$^{1,2}$, Ievgen Vovk$^3$}

\affiliation{
$^1$Universit\'e Paris Cit\'e, CNRS, Astroparticule et Cosmologie, 75006 Paris, France\\
$^2$Laboratory of Astrophysics, \'Ecole Polytechnique F\'ed\'erale de Lausanne, 1015 Lausanne, Switzerland\\
$^3$ Institute for Cosmic Ray Research, The University of Tokyo, 5-1-5 Kashiwa-no-Ha, Kashiwa City, Chiba, 277-8582, Japan 
}

\begin{abstract}
 We notice that Galactic sources of PeV gamma-rays are generally expected to be surrounded by multi-degree-scale extended emission produced as a result of injection of electron-positron pairs in the interstellar medium from interactions of PeV gamma-rays with microwave photons. The surface brightness of such emission may be variable in time because of variability of the PeV emission power of its parent source. A flare of the parent source produces an ``echo'' spreading across the extended emission region that may be detectable on the time scales from decades to millennia. We consider a possibility that the very extended PeV source in the Cygnus region is the produced by the Cygnus X-3 PeV gamma-ray source, possibly being an echo of its past flare.
\end{abstract}
\maketitle


\section{Introduction}

LHAASO telescope has recently revealed a range of astronomical sources of PeV \gr s  \cite{LHAASO:2021gok}. The PeV sky hosts a set of isolated sources powered by cosmic PeVatrons (objects operating particle accelerators boosting particle energies beyond PeV) and possibly regions of extended emission that are difficult to attribute to a specific source because of their large extent on the sky. This is the case for a source of PeV \gr s in Cygnus region of the Galaxy \cite{LHAASO:2023uhj}. Part of the PeV signal from this source is associated to an X-ray binary Cyg X-3 \cite{LHAASO:2025ysm}, while a larger fraction of the PeV flux originates from a very extended source, with photon hits spread over a region of the radius $R\sim 10^\circ$. It is not clear a-priori if this very extended PeV source is related to Cyg X-3 at the distance $D_{CygX3}\simeq 9$~kpc \cite{Reid:2023ksq} or is linked to a much closer Cygnus X star-forming region that is conventionally associated with the ``cocoon'' of enhanced \gr\ flux observed at lower energies by Fermi/LAT, HAWC and LHAASO telescopes \cite{2011Sci...334.1103A,LHAASO:2023uhj,Abeysekara:2021yum} at $D_{cocoon}\simeq 1.5$~kpc distance \cite{2020MNRAS.495.1209R}. 

Cyg X-3 is one of the Galactic microquasars -- binary systems with compact objects that eject jets visible from radio to ultra-high-energy \gr s \cite{LHAASO:2024psv}. It is known to experience flaring activity in \gr s visible in GeV band by Fermi/LAT telescope \cite{2009Sci...326.1512F} and in PeV range by LHAASO \cite{LHAASO:2025ysm}, but not observable in TeV band by Cherenkov telescopes \cite{Abe:2026ieo}. The source includes a compact object accreting from the wind of a Wolf-Rayet star with the \gr\ flares accompanying the episodes of ejections into a jet of the compact object. Most of the power of the source, close to the Eddington luminosity level  $L_{edd}\sim 10^{39} (M / 10~M_{\odot})$~erg/s, is released in the X-ray band \cite{1967ApJ...148L.119G}, while its present-day \gr\ luminosity, $\sim 10^{33}$~erg/s, is much lower. It is possible that the source has experienced a major activity episode half-a-century ago, with substantially increased high-energy flux \cite{1988PhR...170..325B}. 
Claims of detection of $E>500$~TeV \gr s from the source have been published in the 1970th \cite{1986ApJ...301..230K,1987NCimC..10..151A,1986ApJ...306..587B,1988PhR...170..325B}, with the flux measurements in the initial reports of the order of $F\sim 10^{-10}-10^{-9}$~TeV/(cm$^2$s).   

PeV \gr s are known to have limited propagation range, because of pair production on low-energy photons of Cosmic Microwave Background (CMB) radiation \cite{Gould:1967zzb}. This range is comparable to the size of a Milky Way like galaxy and hence all the detected PeV \gr\ sources are situated inside the Milky Way, at most at $\sim 10$~kpc distance. The same process also results in injection of electrons and positrons in the medium all across the  volume of the size comparable \gr\ mean free path. \gr s emitted by such electrons and positrons can potentially be observed by \gr\ telescopes as  halos surrounding the parent source. Such a possibility was considered in Ref. \cite{Aharonian:1993vz} for much larger scale (hundreds of Megaparsec-scale) halos around  extragalactic sources of TeV \gr s. The halos around PeV sources are much more compact (10~kpc scale) and should be observable around sources in the Milky Way galaxy.  We discuss a possibility that the two PeV sources in Cygnus region -- the Cyg X-3 and the extended source -- are related, with the extended source being a trace of the PeV \gr s spreading around Cyg X-3 and interacting with the CMB.

\section{Gamma-ray ``echo'' signal of the PeV flare}

The mean free path of \gr s with energies close to a PeV  is  $D_{\gamma_0}=(\sigma_{\gamma\gamma}n_{CMB})^{-1}\simeq 8$~kpc,
with $\sigma_{\gamma\gamma}\simeq 10^{-25}$~cm$^2$ being the pair production cross-section \cite{Gould:1967zzb}  and $n_{CMB}\simeq 4\times 10^2$~cm$^{-3}$ the number density of the CMB photons. The pair production by the PeV \gr s  results in injection of electrons and positrons in the interstellar medium in the region of the size $\sim D_{\gamma_0}$ around the primary source of PeV \gr s. These electrons and positrons re-emit \gr s with energies
$E_\gamma\simeq \epsilon_{CMB}\left(E_e/m_ec^2\right)^2 \simeq  10^{15}\left[E_e/10^{15}\mbox{ eV}\right]^2\mbox{ eV}$ ($\epsilon_{CMB}\sim 10^{-3}$~eV 
is average energy of CMB photons) through inverse Compton scattering. Electrons and positrons with energies above $\sim 10^{15}$~eV produce the inverse Compton photons in the Klein-Nishina regime in which the cross-section drops below the Thompson cross-section, $\sigma_T$. The cooling distance of the PeV electrons and positrons is somewhat shorter than that of the primary \gr\ mean free path (cooling distance), $D_{e, PeV}=(\sigma_Tn_{CMB})^{-1}\simeq 3-5$~kpc \cite{Lee:1996fp}. Lower energy electrons produce inverse Compton photons in Thompson regime in which the cooling distance scales inversely proportionally to the electron energy, $D_e(E_e)\sim D_{e,PeV}\left[E_e/1\mbox{ PeV}\right]^{-1}$. 

During their cooling time, these electrons and positrons are deflected by $B\sim 1\ \mu\mbox{G}$ magnetic fields present in the interstellar medium. The gyro-radius of the PeV particles is $R_L\sim 10^{16}\left(B/1\ \mu\mbox{G}\right)\left(E_e/1\mbox{ PeV}\right)^2$~cm $\ll D_e$ which means that particle trajectories are deflected by the magnetic field.  Electrons and positrons spiral along regular magnetic field lines and are scattered by inhomogeneities of the turbulent component of the field. This process randomizes electron and positron directions and the secondary inverse Compton photons are most probably  emitted nearly isotropically. We adopt such an assumption in the following estimates. 

The secondary signal reaches the observer with a time delay caused by the finite electron-positron cooling time and also by the geometrical time delay: the secondary photons do not arrive from the direction toward the source. The typical time delay is comparable to the light travel time from the source, $D_s/c\sim 3\times 10^4$~yr for the sources in the Milky Way at the distances $D_s\sim 10$~kpc. The extended emission halo is distributed all over the sky with secondary photons from the direction opposite to the source coming from primary \gr s which initially were emitted almost in the direction of the Earth, passed near the Earth and interacted at the distance larger than the distance from the source to the Earth. 

\section{PeV ``echo'' signal modeling}

To study the timing and imaging properties of the secondary extended emission from electrons and positrons in the interstellar medium, we performed numerical simulations of the signal using the CRpropa code \cite{Aerdker:2023tlu}. We simulated electromagnetic cascade initiated by primary \gr s with energy 3~PeV (close to the maximal photon energy LHAASO registered from the Cyg X-3 direction) emitted isotropically by a point source at the distance of Cyg X-3. We consider two possible simulation setups. First is a stationary source that continuously injects the $3$~PeV \gr s with constant power. Its alternative is a flaring source, for which we assume that the injection of the primary \gr s happens instantaneously.

We follow trajectories of the primary and secondary particles in the electromagnetic cascade within the sphere of the radius $D_s$ and notice the time instants when the primary and/or secondary \gr s in the cascade cross the sphere in the outward direction. For each \gr, we measure the angle between the \gr\ direction and the normal to the sphere, to calculate the angular distribution of the arrival directions of the \gr s forming an extended halo around the primary source position. With such an approach, we miss a part of the overall secondary signal. Most important missed part is the signal coming in the off-source angular range interval  $\theta>90^\circ$, which does not appear because we stop following particle trajectories as soon as they go outside the sphere of the radius $D_s$. We also miss a part of the signal in the $\theta<90^\circ$ interval, because it is also possible that the secondary \gr s that were produced at the distance $D>D_s$ re-enter the sphere $D<D_s$ and then cross the sphere from inside out again. 

The contribution of such re-crossing secondary \gr s to the overall halo signal may be important for nearby sources for which $D_s\ll D_{\gamma_0}$. Halos around such sources are difficult to detect because they are nearly isotropic and their signal is difficult to distinguish from the overall diffuse sky flux. As we consider sources at distances $D_s\gtrsim D_{\gamma_0}$, the contribution of the re-crossing \gr s to the halo flux at moderate off-source angles ($\theta\lesssim 10^\circ$) is minor. 

\section{Results}

\begin{figure}
    \includegraphics[width=\columnwidth]{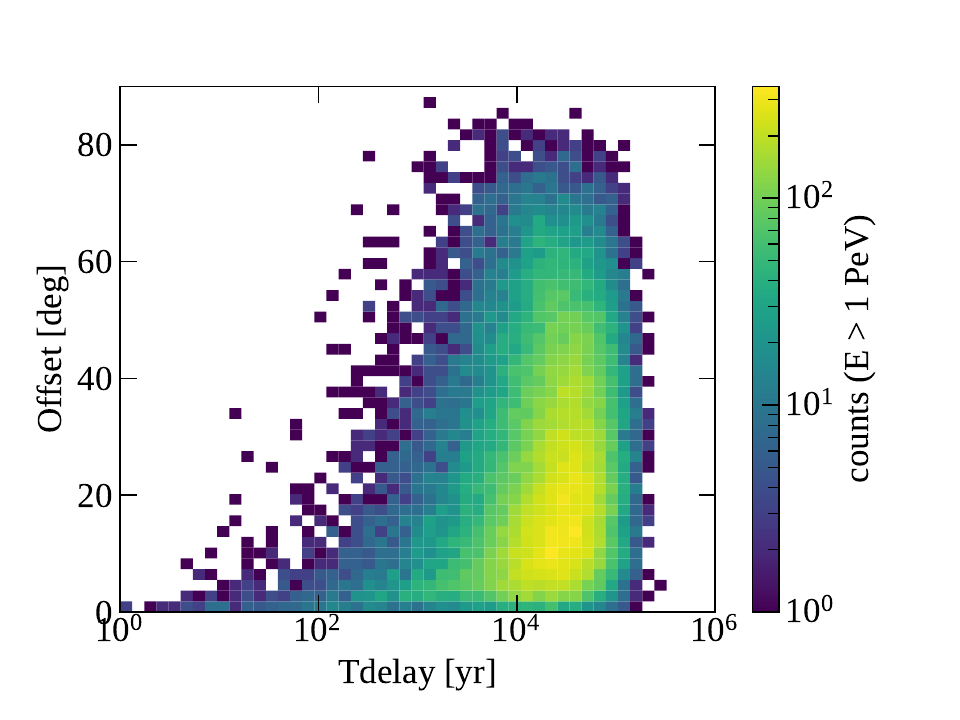}
    \caption{Secondary photon observed offsets as function of their arrival on the time delay. The distribution is plotted for the photon energies above 1~PeV.}
    \label{fig:flux_angle}
\end{figure}

The result of our modeling of a flaring source  is shown in Fig. \ref{fig:flux_angle}, where a distribution of the secondary photon counts is shown as a function of the time delay $T_{delay}$ (after the moment of arrival of the primary $3$~PeV primary \gr s from the source) and the off-source angle $\theta$. One can see that the bulk of the secondary signal indeed arrives with time delay $10^4-10^5$~yr, comparable to the light travel time from the source. As expected, the bulk of the secondary photons is distributed over a very broad angular range, $0<\theta<90^\circ$. 

\begin{figure}
    \includegraphics[width=\columnwidth]{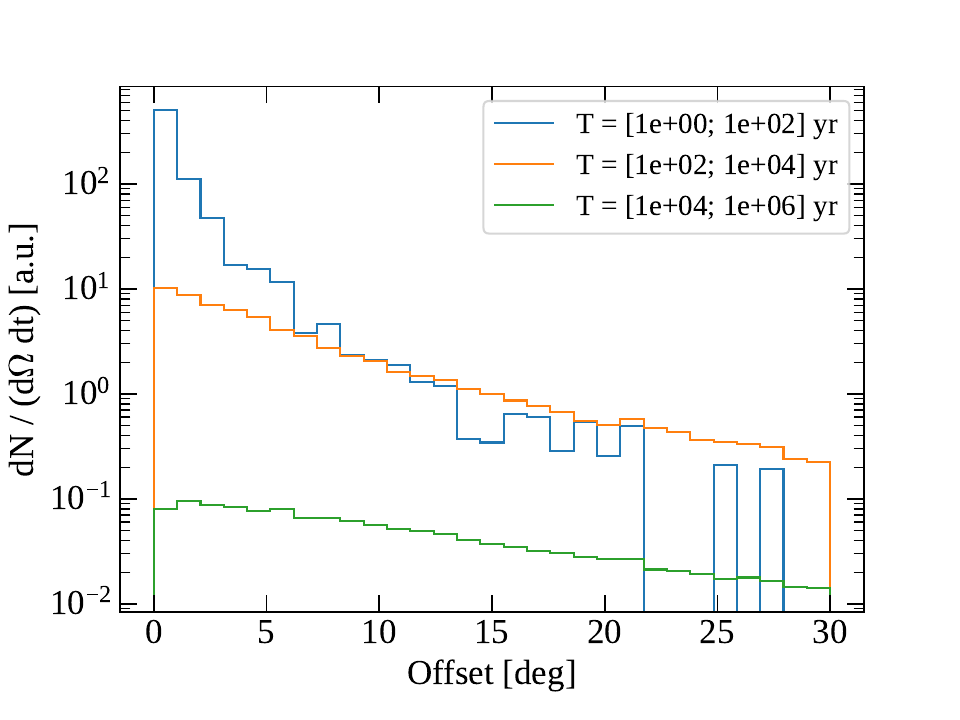}
    \caption{Surface brightness profiles of the secondary extended emission in the energy range above 1~PeV for different time delay ranges.}
    \label{fig:angular_profiles}
\end{figure}

The secondary \gr s that arrive with delays $T_{delay}\ll D_s/c$ are concentrated in a more compact halo around the source. From Fig.~\ref{fig:flux_angle} one can see that photons arriving within the fist $T_{delay}\lesssim 100$~yr are mostly concentrated within the $\theta\lesssim 10^\circ$ angular interval. The surface brightness profiles of the extended emission for different time delay ranges are shown in Fig.~\ref{fig:angular_profiles}. In all cases, the surface brightness of the halo decreases with the angular offset from the source. This means that the halo signal may still be distinguished from other unrelated diffuse sources on the sky: the halo signal is concentrated towards the parent source direction. However, in the case of a very long time delay, the surface brightness contrast decreases: at $30^\circ$ off-source angle the surface brightness is lower than that in the source direction tenfold. To the contrary, during the first 100 years after the flare, the secondary flux is mostly concentrated within $\theta\sim 5^\circ-10^\circ$ region around the source. 

Note that in spite of the much lower photon fluence (the overall energy contained in the signal) evident from Fig. \ref{fig:flux_angle}, the flux of the secondary halo within the $\theta<10^\circ$ region around the source during the first hundred(s) of years after the primary source flare is highest.  We estimate that a fraction $\kappa_{100yr,10^\circ}=1.4\times 10^{-3}$ of the primary source flare energy is re-emitted in this time and angle range in our model setup -- with a much smaller  $\kappa_{100yr,10^\circ-90^\circ}=2.5\times 10^{-4}$ part of it emitted in the $10^\circ-90^\circ$ angular range.

\begin{figure}
    \includegraphics[width=\columnwidth]{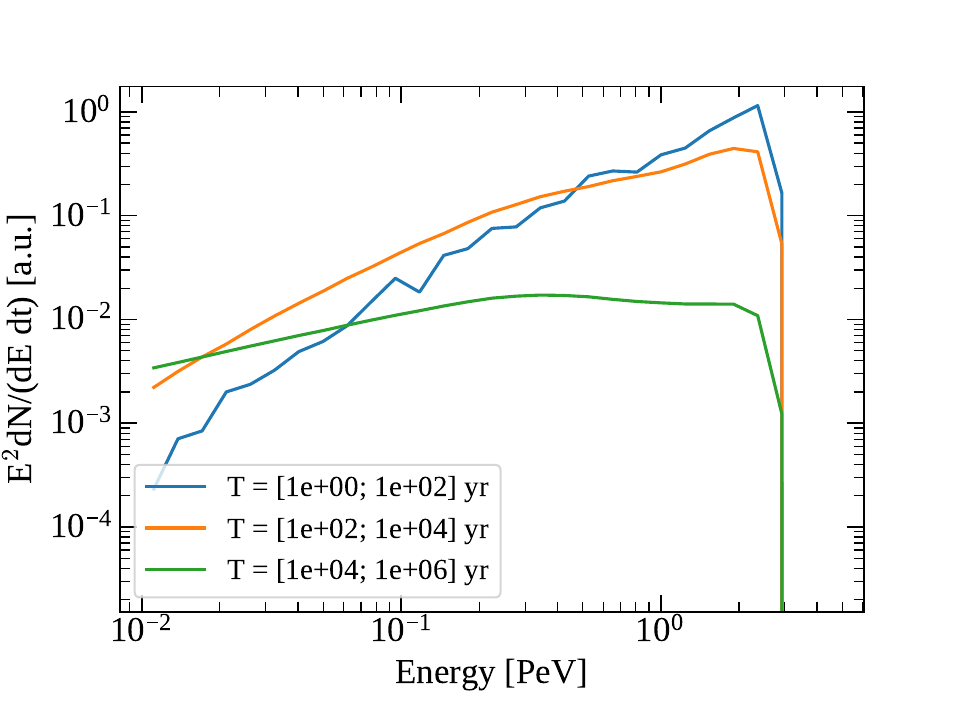}
    \caption{Spectrum of the secondary signal as a function of the time delay. }
    \label{fig:spectrum_tdelay}
\end{figure}

Temporal evolution of the secondary signal spectral energy distribution is given in Fig.~\ref{fig:spectrum_tdelay}. One can see that the initial 100~yr long time interval is characterized by a hard source spectrum -- so that most of the flux forming the compact $5^\circ-10^\circ$ scale halo is concentrated in the PeV energy range. The spectrum in the sub-PeV energy interval follows a power law shape, $dN/dE\propto E^{-\Gamma}$, with very hard slope $\Gamma\simeq 0.5$. The spectrum softens at later times, reaching nearly $\Gamma = 2$ for the bulk of the signal coming within time delay $10^4\mbox{ yr}<T_{delay}<10^6$~yr. The low-energy tail is formed by electrons cooling through the emission of the inverse Compton \gr s. 

\begin{figure}
    \includegraphics[width=\columnwidth]{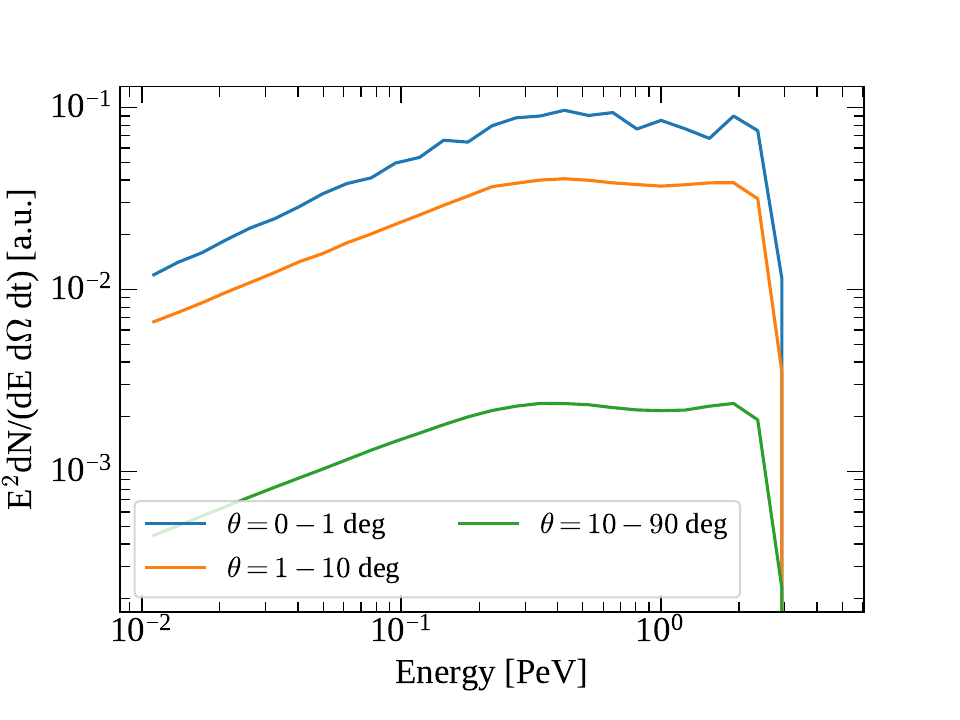}
    \caption{Spectrum of the secondary signal as a function of the off-source angle for a stationary source. }
    \label{fig:spectrum_offset}
\end{figure}

If the source has been a steady PeV \gr\ emitter over the last $10^4-10^5$~yr, it is expected to be surrounded by a very broad halo, with a mild surface brightness contrast of a factor $\simeq 50$ between the emission coming within one-degree region around the source and the broad $90^\circ$ scale halo. This can be seen from Fig.~\ref{fig:spectrum_offset} where the spectra collected from concentric rings $\theta<1^\circ$, $\theta = [1^\circ;10^\circ]$ and $\theta>10^\circ$ are shown. One can see that the halo spectrum is independent of the offset angle, remaining a $dN/dE\propto E^{-2}$ power law in the energy range above $100$~TeV and hardening below that energy. In spite of the larger surface brightness, the more compact part of the halo contains only a minor fraction of the overall secondary emission flux: while a total of $\kappa_{10^\circ}\simeq 0.08$ of the primary source flux is concentrated within $10^\circ$ from the source, as much as $\kappa_{10^\circ-90^\circ}\simeq 0.3$ of it arrives at offsets $\theta>10^\circ$. Observation of extended ten-degree-scale emission around an isolated PeVatron this way provides a measurement of its time-averaged emission power.

\section{Gamma-ray halo around Cyg X-3?}

The model of extended emission around a source of PeV \gr s can be applied to the known Galactic sources of PeV \gr s discovered by LHAASO \cite{LHAASO:2021gok}. All these sources are bound to be surrounded by such halos, whose flux levels can be well predicted if the source emission power history is known (or if they are know to be steady on sufficiently long timescales). This may be the case, e.g. for Crab Nebula, where the luminosity evolution over the entire $\sim 10^3$~yr long source life time span can be reconstructed. However, even for Crab Nebula it is difficult to estimate the PeV energy range emission efficiency throughout its entire history. Similar to the Crab Nebula, pulsar spin down power may also be responsible for some other brightest PeV sources from the list of Ref.~\cite{LHAASO:2021gok}, with some notable exceptions, the sources that may be powered by microquasars \cite{LHAASO:2024psv}.

One such microquasar is located inside a remarkable, $\theta\sim 10^\circ$ scale PeV \gr\ source revealed through the LHAASO study of the Cygnus X region~\cite{LHAASO:2023uhj, LHAASO:2025ysm}.
Out of seven PeV photons in the region, five come from the extended source, with other two spatially coincident with the position of Cyg X-3 binary system. Contrary to young pulsars, this binary system is a highly variable source and its past activity history is largely uncertain. Earlier claims~\cite{1986ApJ...301..230K,1987NCimC..10..151A,1986ApJ...306..587B,1988PhR...170..325B} suggest the source detection at energies $E>500$~TeV with the flux reaching $F\sim 10^{-10}-10^{-9}$~TeV/(cm$^2$s) -- several orders of magnitude higher compared to the present-day PeV-band flux level \cite{LHAASO:2025ysm}. It thus remains plausible that the source experienced major flaring activity few decades ago and subsequently entered its present-day quieter state. As the modern LHAASO measurements, in general, may be not representative of the average source luminosity on $10^5$~yr time scale, the duty cycle of such enhanced activity periods remains uncertain.


The steady-state halo model allows to estimate the flux within a $10^\circ$ scale region around the source to be $\kappa_{10^\circ}\sim 0.1$ of the time-averaged primary source flux over the $\sim 10^5$~yr time scale. LHAASO detection of five PeV photons from the extended source compared to only two events from Cyg X itself suggests halo flux is 2-3 times larger than that of the point source seen presently. This itself implies that Cyg X-3 nowadays activity is particularly low with the PeV flux at the level of 20-30 times lower than the hundred-thousand-year average level. Worthy to note that this 20-30 times higher flux is still well below the Eddington luminosity and is lower than the measured X-ray flux of the source.


It is equally plausible possibility that the extended emission seen by LHAASO is an ``echo'' recent flaring activity of the source. As discussed above, during the first 100~years after the flare the echo signal remains concentrated within the $10^\circ$ region (which would be the case the flaring activity reported several decades ago). Our estimates suggest that the PeV band fluence of the extended emission is of the order of $\kappa_{100yr,10^\circ}\sim 10^{-3}$ of the total flare energy output. Assuming the flare duration of $T_{flare}=10$~yr, one can thus estimate source luminosity to have been $L_{flare}=E_{flare}/T_{flare}\sim 10^{37}$~erg/s. Such a flare would yield the flux of $F_{flare}\sim 10^{-9}$~TeV/(cm$^2$s) -- compatible with that reported in the PeV band in 1970th within an order-of-magnitude.


\section{Summary and conclusions}

We have shown that Galactic sources of PeV \gr s recently discovered by LHAASO are generically expected to be surrounded by multi-degree-scale extended emission ``halos'' that may have time-dependent morphology. The very extended PeV \gr\ source in the Cygnus region of the Galaxy may actually be such an extended halo around the binary system Cyg X-3. Our numerical modeling and back-of-the-envelope estimates show that the energetics and morphology of the extended source are consistent with a possibility that it is an ``echo'' of an activity episode of Cyg X-3 that occurred half-a-century ago. 

\bibliography{refs}
\end{document}